\documentclass[%
 reprint,
 onecolumn,
 amsmath,amssymb,
 aps,
]{revtex4-2}

\setcitestyle{super}
\usepackage{graphicx}
\usepackage{dcolumn}
\usepackage{bm}
\usepackage{hyperref}
\usepackage{color}

\begin{document}


\title{Local temperature measurement in molecular dynamics simulations with rigid constraints}

\author{Stephen Sanderson}
\email{stephen.sanderson@uq.edu.au}
 \affiliation{Australian Institute for Bioengineering and Nanotechnology, The University of Queensland, St. Lucia, QLD, 4072, Australia}
\author{Shern R. Tee}
 \affiliation{Australian Institute for Bioengineering and Nanotechnology, The University of Queensland, St. Lucia, QLD, 4072, Australia}
 \affiliation{School of Environment and Science - Chemistry and Forensic Science, Griffith University, Nathan, QLD, 4111, Australia}
\author{Debra J. Searles}
 \email{d.bernhardt@uq.edu.au}
 \affiliation{Australian Institute for Bioengineering and Nanotechnology, The University of Queensland, St. Lucia, QLD, 4072, Australia}
 \affiliation{School of Chemistry and Molecular Biosciences, The University of Queensland, St. Lucia, QLD, 4072, Australia }
  \affiliation{ARC Centre of Excellence for Green Electrochemical Transformation of Carbon Dioxide, The University of Queensland, St. Lucia, QLD, 4072, Australia }

\date{\today}

\begin{abstract}
Constraining molecules in simulations (such as with constant bond lengths and/or angles) reduces their degrees of freedom (DoF), which in turn affects temperature calculations in those simulations.
When local temperatures are measured, e.g. from a set of atoms in a subvolume or from velocities in one Cartesian direction, the result can appear to unphysically violate equipartition of the kinetic energy if the local DoF are not correctly calculated.
Here we determine how to correctly calculate local temperatures from arbitrary Cartesian component kinetic energies, accounting for general geometric constraints, by self-consistently evaluating the DoF of atoms subjected to those constraints.
The method is validated on a variety of test systems, including systems subject to a temperature gradient and those confined between walls.  
It is also shown to provide a sensitive test for the breakdown of kinetic energy equipartition caused by the approximate nature of numerical integration or insufficient equilibration times.
As a practical demonstration, we show that kinetic energy equipartition between C and H atoms connected by rigid bonds can be violated even at the commonly-used time step of 2 fs, and that this equipartition violation appears to usefully indicate configurational overheating.

\end{abstract}

\maketitle


\section{Introduction}

In molecular dynamics (MD) simulations, nanoscale temperature profiles are often required.
For example, temperature profiles in confined fluid flows \cite{Baranyai1992,Todd1997} or across solid-liquid interfacial discontinuities \cite{Alosious2024,Muscatello2017,Alosious2020,Hu2012} are required to gain molecular insights into thermal conductivity and local heating.
In such simulations, locally-averaged particle thermal kinetic energies provide the local temperature, or more strictly the local ``kinetic temperature''.
``Cartesian'' kinetic temperatures (components of the kinetic temperature due to velocities in the $x$, $y$ and $z$ directions)  are also useful for correctly thermostatting non-equilibrium simulations.
For example, calculating the temperature of a group of atoms in a fluid that is flowing requires subtraction of the streaming, non-thermal velocity, but using only transverse velocity components is an expedient alternative \cite{Toton2010}.

MD simulations also often impose rigid constraints using constraint dynamics \cite{Edberg1986} or algorithms such as SHAKE \cite{Ryckaert1977}, RATTLE \cite{Andersen1983} or LINCS \cite{Hess1997}.
The temperature of a system of atoms with constraints is straightforwardly calculated by subtracting one degree of freedom (DoF) for each constraint imposed.
However, when a local temperature calculation includes some, but not all, atoms or Cartesian directions which participate in a given constraint, it is not straightforward to determine how many DoF should be associated with that subset.
Popular molecular dynamics codes often simply split DoF evenly between atoms (such as treating two atoms joined by a bond of fixed length by subtracting half a DoF each), and this approach often gives an approximately correct result on average, but any inhomogeneity (such as an interface or alignment under a field) can result in apparent differences in temperature between neighbouring subsets, even at equilibrium where the equipartition theorem dictates that their average temperatures should match \cite{Tolman1927}.
This problem could be further amplified when a thermostat is applied to the (incorrect) local temperature, potentially resulting in the simulation having a vastly different temperature to the one prescribed.
Such problems have been identified, for example, in studies of the liquid-solid interface \cite{OlartePlata2022} and polar molecules under an electric field \cite{Gullbrekken2023}.

Recently this issue has been considered for some specific cases,\cite{OlartePlata2022,Matsubara2023}, but as highlighted by Matsubara et al.\cite{Matsubara2023}, the approaches presented to date are not easily extended into a practical, generally applicable procedure.
For example, treatment of systems with many connected constraints, such as a long carbon chain or a PF$_6^-$ ion with rigid bonds, would be challenging and require new derivations.
Below we present an approach where these extensions are straightforward and, in addition, components of the kinetic temperature in different directions are directly obtained. 
The results presented in this manuscript can be applied to any dynamical particle systems with geometric constraints, although the concrete examples focus on molecular systems since geometrically-constrained molecular dynamics simulations are widely applied to systems as disparate as biological macromolecules and energy storage materials.
Thus ``molecules'' could be more generally interpreted as sets of particles which may have constraints between them, and ``atoms'' the individual particles (so, for example, they could be united atoms or other coarse-grained point masses).

\section{Theory}

Suppose we have a system of molecules consisting of the set of atoms, $\mathcal{P}$, moving in $\mathcal{D}$ dimensions (usually $\{x,y,z\}$), and we are interested in a subset of atoms, $\mathcal{S}$, which each have mass $m_j$ and peculiar (thermal) velocity $\mathbf{v}_j$.
The local kinetic temperature of that subset, $T_\mathcal{S}$, is given by
\begin{equation}
    T_\mathcal{S} = \frac{1}{2d_\mathcal{S}k_B}\left<\sum_{j\in\mathcal{S}}\sum_{\alpha\in \mathcal{D}} m_j\left(\mathbf{v}_j\cdot\mathbf{\hat{e}}_\alpha\right)^2\right>, \label{eqn:local-temperature}
\end{equation}
where $k_B$ is Boltzmann's constant, $\langle\dots\rangle$ represents an ensemble average, $\hat{\mathbf{e}}_\alpha$ is the unit vector in direction $\alpha$, and $d_\mathcal{S}$ is the number of degrees of freedom in the subset.

In the absence of constraints, $d_\mathcal{S}$ is simply $\left|\mathcal{D}\right|\left|\mathcal{S}\right|$; that is, the number of spatial dimensions multiplied by the number of atoms in the subset ($d_\mathcal{S} = 3N_\mathcal{S}$ for a system of $N_\mathcal{S}\equiv\left|\mathcal{S}\right|$ atoms in three dimensions).
If a set of $N_c$ constraints acts on the relative motions of particles which are all in $\mathcal{S}$, then one degree of freedom should be subtracted for each such constraint.
That is, $d_\mathcal{S}$ in Equation \ref{eqn:local-temperature} becomes $|\mathcal{D}||\mathcal{S}| - N_c$.
For example, if we consider CH$_3$CCl$_3$ in 3 Cartesian dimensions, and constrain the 3 C-H bond lengths, the number of degrees of freedom of the subset of atoms in CH$_3$ will be $d_\mathcal{S}=(3 \times 4 - 3)$ whereas for the subset in CCl$_3$ it will be $d_\mathcal{S}=(3 \times 4)$ and the total number of degrees of freedom of the molecule will be $d=(6 \times 4 -3)$.

\begin{figure}
    \centering
    \includegraphics[width=0.5\linewidth]{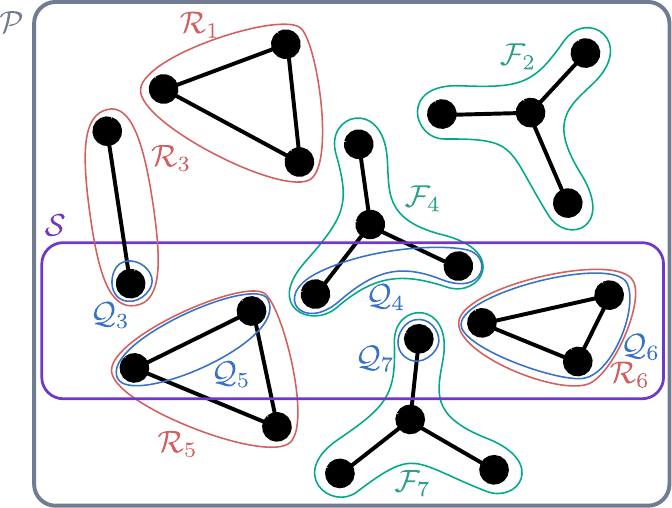}
    \caption{%
        An illustration of the system, constrained atoms, and various subsets relevant to local temperature measurement.
        $\mathcal{P}$ is the set of all atoms in the system.
        $\mathcal{S}$ is the subset of interest for local temperature measurement $(\mathcal{S} \subseteq\mathcal{P})$.
        $\mathcal{R}_i$ are sets of atoms which form rigid bodies ($\mathcal{R}_i\subseteq\mathcal{P}$).
        $\mathcal{F}_i$ are sets of atoms which form semi-rigid fragments ($\mathcal{F}_i\subseteq\mathcal{P}$).
        $\mathcal{Q}_i$ are the subsets of $\mathcal{R}_i$ and $\mathcal{F}_i$ which intersect with $\mathcal{S}$.
        The DoF associated with a given $\mathcal{Q}_i$ is straightforward to determine when all atoms in a constraint are included, and thus $\mathcal{Q}_i = \mathcal{R}_i$ or $\mathcal{Q}_i = \mathcal{F}_i$ (e.g. $\mathcal{Q}_6$), while the methods described in this manuscript are needed for correct DoF determination when some constrained atoms are excluded from the local temperature measurement, making $\mathcal{Q}_i$ a proper subset of $\mathcal{R}_i$ or $\mathcal{F}_i$, so that $\mathcal{Q}_i \neq \mathcal{R}_i$ or $\mathcal{Q}_i \neq \mathcal{F}_i$ (e.g. $\mathcal{Q}_3$, $\mathcal{Q}_4$, $\mathcal{Q}_5$ or $\mathcal{Q}_7$).
    }
    \label{fig:sets}
\end{figure}

However, if we now defined $\mathcal{S}$ as just the H atoms, it is no longer straightforward to determine the correct $d_\mathcal{S}$ to use, since the set of constraints involves atoms both within and outside the subset of interest.
This work aims to treat such cases, which are further illustrated in Fig. \ref{fig:sets}. 

\subsection{Rigid Bodies\label{sec:theory:rigid}}
We first consider the case of a fully rigid body, $i$, that could be a molecule or a molecular fragment. The rigid body comprises an arbitrary set of atoms, $\mathcal{R}_i$, in 3-dimensional space ($\mathcal{D} = \left\{x, y, z\right\}$) and its kinetic energy is given by
\begin{eqnarray}
    K_i &=& \frac{1}{2}\sum_{j\in\mathcal{R}_i} m_j\mathbf{v}_j\cdot\mathbf{v}_j, \\
                  &=& \frac{1}{2}\left(\mathbf{v}_i^T\mathbf{M}_i\mathbf{v}_i
                      + \boldsymbol{\omega}_i^T\mathbf{I}_i\boldsymbol{\omega}_i \right), \\
                  &=& K_{\mathcal{R}_i}^\text{tr} + K_{\mathcal{R}_i}^\text{rot},
\end{eqnarray}
where $\mathbf{v}_i$ is the center of mass velocity of the rigid body, $\boldsymbol{\omega}_i$ is its angular velocity,
\begin{equation}
    \mathbf{M}_i = \sum_{j\in\mathcal{R}_i}m_j\mathbf{1}
\end{equation}
is its translational inertia matrix (or mass matrix), with $\mathbf{1}$ being the identity/unit matrix of size 3, and
\begin{eqnarray}
    \mathbf{I}_i &=& \sum_{j\in\mathcal{R}_i} -m_j\left[\begin{array}{ccc}
        0 & -r_{ij,z} & r_{ij,y} \\
        r_{ij,z} & 0 & -r_{ij,x} \\
        -r_{ij,y} & r_{ij,x} & 0
    \end{array}\right]^2 \\
                &\equiv& \sum_{j\in\mathcal{R}_i} -m_j \left[\mathbf{r}_{ij,\times}\right]^2 \label{eqn:rot-inertia}\\
                &=& \sum_{j\in\mathcal{R}_i} \mathbf{I}_{ij}
\end{eqnarray}
is its rotational inertia tensor (or moment of inertia tensor) relative to the center of mass of the rigid body.
Here $r_{ij,\alpha}$ is the $\alpha$ component of the position of atom $j$ relative to the
center of mass of its rigid body, $i$,  and the $\times$ subscript notation, defined by Eqn. \ref{eqn:rot-inertia}, denotes a matrix which applies the cross product; that is, $[\mathbf{r}_{ij,\times}]\mathbf{v} \equiv \mathbf{r}_{ij} \times \mathbf{v}$ for a vector $\mathbf{v}$.

We start by considering the process for treating the translational DoF associated with each mode (i.e. translation in the $x$, $y$ and $z$ directions).
For self-consistency it is required that at equilibrium the DoF of each mode  must be partitioned between atoms of the rigid body such that the same temperature is obtained when it is calculated for any non-empty subset of atoms, $\mathcal{Q}_i$, within $\mathcal{R}_i$.
That is, for direction $\alpha$,
\begin{equation}
    \frac{\left< K_{\mathcal{Q}_i}^{\text{tr},\alpha}\right>}{d_{\mathcal{Q}_i}^{\text{tr},\alpha}} =
    \frac{\left< K_{\mathcal{R}_i}^{\text{tr},\alpha}\right>}{d_{\mathcal{R}_i}^{\text{tr},\alpha}} \hspace{2em}\forall \mathcal{Q}_i \subseteq \mathcal{R}_i,
\end{equation}
where $K_{\mathcal{R}_i}^{\text{tr},\alpha}$ is the translational kinetic energy along direction $\alpha$, so
$K_{\mathcal{R}_i}^\text{tr} = \sum_{\alpha\in D} K_{\mathcal{R}_i}^{\text{tr},\alpha}$.
Since the translational velocity of each atom of the rigid body must be the same and the subset $\mathcal{Q}_i$ could be the single atom $j$, this leads to a unique solution for the DoF of atom $j$ in translational mode $\alpha$ of rigid body $i$ as
\begin{eqnarray}
    d_{j}^{\text{tr},\alpha} &=& \frac{m_j}{\sum_{k\in\mathcal{R}_i}m_k}, \\
                           &=& \frac{\mathbf{\hat{e}}_\alpha^T\left(m_j\mathbf{1}\right)\mathbf{\hat{e}}_\alpha}
                                    {\mathbf{\hat{e}}_\alpha^T\mathbf{M}_i\mathbf{\hat{e}}_\alpha}.
\label{eqn:trdof}
\end{eqnarray}
The first line shows that this is a simple partitioning based on atom $j$'s contribution to the total mass of the rigid body.
However, we write it in a more general form in the second line to aid generalisation later.
Hence, the translational DoF should be partitioned based on the contribution of each atom participating in the mode to the translational inertia (mass). 
In the example considered above, if the CH$_3$ group is completely rigid, all atoms in this group will have the same translational velocity and, according to Eqn. \ref{eqn:trdof}, the C atom has $d_C^{tr}=3 \times 12.011/15.035$ and each H atom has $d_H^{tr}=3 \times 1.008/15.035$, with a total of $d^{tr}=3$ for the CH$_3$ group.
For the translation degrees of freedom, the partitioning only depends on the masses of the atoms and not on the structure of the rigid molecule or molecular fragment.
We note in passing that this partitioning also applies to the DoF reduction associated with the constraint on the total center of mass momentum for a system with only internal forces, as has been demonstrated by Uline \textit{et al.} \cite{Corti2008}.

This process extends straightforwardly to the rotational modes.
The rotational inertia tensor, $\mathbf{I}_i$, is symmetric positive semi-definite, and thus always has an eigendecomposition $\mathbf{I}_i = \mathbf{Q}_i \mathbf{\Lambda}_i \mathbf{Q}_i^T$, in terms of the projection matrix of eigenvectors $\mathbf{Q}_i$ and the diagonal matrix of eigenvalues $\mathbf{\Lambda}_i$.
As such, the columns $\mathbf{\hat{q}}_{i\beta}$ of $\mathbf{Q}_i$ (representing the normal modes of rotation) are the orthonormal unit vectors pointing in the principal axes, $\beta$, and each diagonal entry, $\lambda_{i\beta}$, of $\mathbf{\Lambda}_i$ is the corresponding principal moment of inertia.
Hence the rotational kinetic energy of any subset $\mathcal{Q}_i$ can be decomposed into components from each normal mode (i.e. about each principal axis) by
\begin{equation}
    K_{\mathcal{Q}_i}^\text{rot} = \sum_{\beta\in\mathcal{D}} K_{\mathcal{Q}_i}^{\text{rot},\beta},
\end{equation}
where
\begin{eqnarray}
    K_{\mathcal{Q}_i}^{\text{rot},\beta} &\equiv&
    \frac{1}{2}\sum_{j\in\mathcal{Q}_i}\boldsymbol{\omega}_i^T\mathbf{I}_{ij}\boldsymbol{\omega}_i,\\
    &=&\frac{1}{2}\sum_{j\in\mathcal{Q}_i}
    \boldsymbol{\omega}_i^T \mathbf{\hat{q}}_{i\beta} \mathbf{\hat{q}}_{i\beta}^T\mathbf{I}_{ij}\mathbf{\hat{q}}_{i\beta} \mathbf{\hat{q}}_{i\beta}^T\boldsymbol{\omega}_i,\\
    &=& \frac{1}{2}\boldsymbol{\omega}_i^T \mathbf{\hat{q}}_{i\beta}  \mathbf{\hat{q}}_{i\beta}^T\boldsymbol{\omega}_i \sum_{j\in\mathcal{Q}_i}
     \mathbf{\hat{q}}_{i\beta}^T\mathbf{I}_{ij}\mathbf{\hat{q}}_{i\beta},\\
     &=& \frac{1}{2}\left|\boldsymbol{\omega}_i\right|^2\sum_{j\in\mathcal{Q}_i}
     \mathbf{\hat{q}}_{i\beta}^T\mathbf{I}_{ij}\mathbf{\hat{q}}_{i\beta}.
\end{eqnarray}
In the first line, $\boldsymbol{\omega}_i$ is projected onto the basis vector of mode $\beta$ in order to determine the kinetic energy in that mode, while the following lines show a convenient simplification, noting that $\left|\hat{\mathbf{q}}_{i\beta}\right|=1$.

To calculate the partitioning of the DoF for each non-zero rotational mode ($\lambda_{i\beta}\neq0$), we again require for self-consistency that any subset $\mathcal{Q}_i \subseteq \mathcal{R}_i$ has equal modal temperature when at least one particle in $\mathcal{Q}_i$ participates in the mode; that is
\begin{equation}
    \frac{\left< K_{\mathcal{Q}_i}^{\text{rot},\beta}\right>}{d_{\mathcal{Q}_i}^{\text{rot},\beta}} =
    \frac{\left< K_{\mathcal{R}_i}^{\text{rot},\beta}\right>}{d_{\mathcal{R}_i}^{\text{rot},\beta}} \hspace{2em}\forall\mathcal{Q}_i\subseteq\mathcal{R}_i, \lambda_{i\beta}\neq0.
\end{equation}
Considering the subset $\mathcal{Q}_i$ to be the single atom $j$ leads to the solution
\begin{equation}
    d_{j}^{\text{rot},\beta} = \frac{\mathbf{\hat{q}}_{i\beta}^T\mathbf{I}_{ij}\mathbf{\hat{q}}_{i\beta}}
             {\mathbf{\hat{q}}_{i\beta}^T\mathbf{I}_{i}\mathbf{\hat{q}}_{i\beta}} = \frac{\mathbf{\hat{q}}_{i\beta}^T\mathbf{I}_{ij}\mathbf{\hat{q}}_{i\beta}}{\lambda_{i\beta}},
             \label{eqn:dof-rot}
\end{equation}
where the denominator is the total  inertia of mode $\beta$.
Hence, as for the translational case, the DoF associated with a given rotational mode is partitioned between atoms based on their contribution to the modal inertia.  However, in this case the DoF will depend on both the mass of the atoms and the geometry of the rigid body through $\mathbf{I}_{ij}$.

The above is sufficient to calculate the total DoF associated with particle $j$ in a rigid body as
\begin{equation}
    d_j = \sum_{\alpha\in\mathcal{D}} d_{j}^{\text{tr},\alpha} + \sum_{\substack{\beta\in\mathcal{D}\\\lambda_{i\beta}\neq0}}d_{j}^{\text{rot},\beta}, \label{eqn:dof-per-atom-total}
\end{equation}
and hence to obtain the local temperature $T_\mathcal{S}$ from Eqn. \ref{eqn:local-temperature} based on
\begin{equation}
    d_\mathcal{S} = \sum_{j\in\mathcal{S}} d_j.
\end{equation}
This is a generalisation of the method presented in Ref. \cite{OlartePlata2022} for rigid water molecules viewed from a modal reference frame, and the result is equivalent to that recently derived by Matsubara \textit{et al.} \cite{Matsubara2023} based on equilibrium statistical mechanics.
Directly applying these equations to the same rigid body geometry of the simple point charge extended (SPC/E) water model \cite{Berendsen1987} yields values of 1.5947 and 2.8106 DoF for H and O atoms, respectively, regardless of the reference coordinate system.
This is in agreement with previously published values \cite{OlartePlata2022,Matsubara2023}, and testing on other geometries confirmed that symmetrically equivalent particles within a rigid body are always assigned equal DoF as expected.
Further validation by simulation of rigid ethane molecules in contact with a flexible graphene membrane can be seen in Section \ref{sec:results:rigid-ethane}.

\subsubsection{Directional temperature}
We now extend this framework to consider cases where the local temperature due to components of the thermal kinetic energy in particular orthogonal directions is required, such as measurement of temperature parallel or perpendicular to an interface.
In these cases, the orthogonal unit vectors in the directions of interest, $\mathbf{\hat{e}}_\gamma$, are not necessarily aligned with the basis vectors of the reference frame, $\mathbf{\hat{e}}_\alpha$.

For translational DoF, it is trivial to see from projection of each mode along each $\mathbf{\hat{e}}_\gamma$ that
\begin{eqnarray}
    d_{j}^{\text{tr},\gamma}
    &=& \sum_{\alpha\in\mathcal{D}} d_{j}^{\text{tr},\alpha\gamma} \\
    &=& \sum_{\alpha\in\mathcal{D}}
        \frac{\mathbf{\hat{e}}_\alpha^T
              \mathbf{\hat{e}}_\gamma\mathbf{\hat{e}}_\gamma^T
              \left(m_j\mathbf{1}\right)
              \mathbf{\hat{e}}_\gamma\mathbf{\hat{e}}_\gamma^T
              \mathbf{\hat{e}}_\alpha}
             {\mathbf{\hat{e}}_\alpha^T
              \mathbf{\hat{e}}_\gamma\mathbf{\hat{e}}_\gamma^T
              \mathbf{M}_i
              \mathbf{\hat{e}}_\gamma\mathbf{\hat{e}}_\gamma^T
              \mathbf{\hat{e}}_\alpha}, \\
      &=& \frac{m_j}{\sum_{k\in\mathcal{R}_i}m_k},
\end{eqnarray}
and therefore, as expected, the partitioning of translational DoF is invariant to rotation of the coordinate system.

To find the DoF of atom $j$ associated with motion in direction $\gamma$ due to rotational mode $\beta$, Eqn. \ref{eqn:dof-rot} can be decomposed in a similar manner to obtain the directional inertia of each mode.
For rotational motion in mode $\beta$ with angular velocity of magnitude $a$, the velocity of atom $j\in\mathcal{R}_i$, projected along direction $\gamma$, is given by
\begin{equation}
    v_{j}^{\text{rot},\beta\gamma} = a\mathbf{\hat{e}}_\gamma^T\left[-\mathbf{r}_{ij,\times}\right]\mathbf{\hat{q}}_{i\beta},
\end{equation}
and hence the inertia of that atom in direction $\gamma$ of mode $\beta$ is
\begin{eqnarray}
    I_j^{\text{rot},\beta\gamma}
        &=& m_j\frac{\left(v_j^{\text{rot},\beta\gamma}\right)^2}{a^2}, \\
        &=& m_j\left(\mathbf{\hat{e}}_\gamma^T\left[-\mathbf{r}_{ij,\times}\right]\mathbf{\hat{q}}_{i\beta}\right)^2.
\end{eqnarray}
As before, the fractional DoF associated with motion in mode $\beta$ of atom $j$ in direction $\gamma$ is obtained from the relative inertia contribution as
\begin{equation}
    d_{j}^{\text{rot},\beta\gamma} = \frac{I_j^{\text{rot},\beta\gamma}}{\lambda_{i\beta}}.
\end{equation}
Summing over all translational and rotational modes yields the total DoF of atom $j$ for motion in direction $\gamma$,
\begin{equation}
    d_j^\gamma = \sum_{\alpha\in\mathcal{D}} d_j^{\text{tr},\alpha\gamma} + \sum_{\substack{\beta\in\mathcal{D}\\\lambda_{i\beta}\neq0}} d_j^{\text{rot},\beta\gamma},
    \label{eqn:dof-directional}
\end{equation}
and hence the directional kinetic temperature of a subset of atoms, $\mathcal{S}$, measured along some set of mutually orthogonal $\mathbf{\hat{e}}_\gamma$ for $\gamma\in \mathcal{G}$ can be obtained from
\begin{equation}
    T_{\mathcal{S},\mathcal{G}} = \frac{\sum_{j\in\mathcal{S}}\sum_{\gamma\in \mathcal{G}} m_j\left(\mathbf{v}_j\cdot\mathbf{\hat{e}}_\gamma\right)^2}{k_B\sum_{j\in\mathcal{S}}\sum_{\gamma\in\mathcal{G}} d_j^\gamma}.
\end{equation}
Note that since motion is treated here in the global reference frame rather than a body-fixed frame, it allows direct calculation of DoF in Cartesian directions without requiring further transformations.
Alternatively, since the directional DoF in a body-fixed frame is rotationally invariant, if it has already been calculated in one reference frame (or orientation) then it may be rotated into another.
We demonstrate and validate this directional DoF decomposition in Section \ref{sec:results:water}.

\subsubsection{Volumetric rigid bodies}
It is interesting to consider that while this formalism has been derived in terms of rigid bodies composed of point masses, it could be equally applied to volumetric rigid bodies described by a mass density function, $\rho_i(\mathbf{r})$.
It is simple to see that the density of DoF for the translational mode in direction $\alpha$ is given by
\begin{equation}
    d_i^{\text{tr},\alpha}(\mathbf{r}) = \frac{\rho_i(\mathbf{r})}{\iiint_{-\infty}^{\infty}d^3\mathbf{r} \rho_i(\mathbf{r})}.
\end{equation}
For the rotational modes, the contribution of point $\mathbf{r}$ to the rotational inertia is given by
\begin{equation}
    \mathbf{I}_i(\mathbf{r}) = -\rho_i(\mathbf{r})\left[(\mathbf{r}-\mathbf{r}_i)_\times\right]^2,
\end{equation}
where $\mathbf{r}_i$ is the center of mass of the rigid body, and the $\times$ subscript is as defined in Eqn. \ref{eqn:rot-inertia}.
The total rotational inertia of the rigid body is then given by
\begin{equation}
    \mathbf{I}_i = \iiint_{-\infty}^{\infty}d^3\mathbf{r} \mathbf{I}_i(\mathbf{r}).
\end{equation}
Hence, with $\mathbf{Q}_i$ defined as before from the eigenvectors of $\mathbf{I}_i$ such that $\mathbf{Q}_i^T\mathbf{I}_i\mathbf{Q}_i$ is diagonal, the density of DoF associated with rotational mode $\beta$ can be obtained as for the translational modes from the fractional contribution of a point in space towards the total inertia of the mode,
\begin{equation}
    d_i^{\text{rot},\beta}(\mathbf{r}) = \frac{\mathbf{\hat{q}}_{i\beta}^T\mathbf{I}_i(\mathbf{r})\mathbf{\hat{q}}_{i\beta}}{\lambda_{i\beta}}.
\end{equation}
Hence, the total density of DoF for rigid body $i$ is
\begin{equation}
    d_i(\mathbf{r}) = \sum_{\alpha\in\mathcal{D}} d_i^{\text{tr},\alpha}(\mathbf{r}) + \sum_{\substack{\beta\in\mathcal{D}\\\lambda_{i\beta}\neq0}} d_i^{\text{rot},\beta}(\mathbf{r}).
\end{equation}
From this, the kinetic temperature in a subvolume, $\mathcal{V}$, of a system containing $N$ rigid bodies can be obtained as
\begin{equation}
    T_{\mathcal{V}} = \frac{\iiint_\mathcal{V}d^3\mathbf{r} \sum_{i=1}^N \rho_i(\mathbf{r})\mathbf{v}_i(\mathbf{r})\cdot\mathbf{v}_i(\mathbf{r})}{k_B\iiint_\mathcal{V}d^3\mathbf{r} \sum_{i=1}^N d_i(\mathbf{r})},
\end{equation}
where $\mathbf{v}_i(\mathbf{r})$ is the velocity induced at position $\mathbf{r}$ due to the combined translational and rotational motion of rigid body $i$.
This treatment reduces to the previous consideration of point mass rigid bodies when $\rho_i(\mathbf{r})$ is a sum of Dirac delta functions.

An intriguing consequence of this formalism is the possibility to observe a temperature gradient across a rigid body in a system undergoing heat flow by measuring kinetic temperature in subvolumes which are relatively small.
This is investigated in Section \ref{sec:results:dumbbells}.

\subsection{Semi-rigid fragments}
Finally, we demonstrate that this framework can be extended in a practical manner to arbitrary sets of constraints by considering an internal coordinate system for semi-rigid fragments such that the constraints are implicit.
For example, again considering CH$_3$CCl$_3$, this time with rigid C-H bond lengths but angles which are free to change, the CH$_3$ atoms form a semi-rigid fragment since they are all connected by at least one rigid constraint.
The kinetic energy of a semi-rigid fragment, $i$, consisting of the set of atoms $j\in\mathcal{F}_i$ is given by \cite{Kneller1994}
\begin{equation}
    K_i = \frac{1}{2} \dot{\boldsymbol{\theta}}_i^T \boldsymbol{\mathcal{I}}_i \dot{\boldsymbol{\theta}}_i = \frac{1}{2}\dot{\boldsymbol{\theta}}_i^T\mathbf{J}_i^T\left[\begin{array}{ccc}
        m_1\mathbf{1} & \cdots & \mathbf{0} \\
        \vdots & \ddots & \vdots \\
        \mathbf{0} & \cdots & m_j\mathbf{1}
    \end{array}\right]\mathbf{J}_i\dot{\boldsymbol{\theta}}_i,
    \label{eqn:semi-rigid-ke}
\end{equation}
where $\dot{\boldsymbol{\theta}}_i$ is the velocity in the internal coordinate system, $\boldsymbol{\mathcal{I}}_i$ is the inertia, and $\mathbf{J}_i$ is the Jacobian with rank equal to the total DoF which transforms from internal coordinate velocities to lab-frame atom velocities 
(further described below).
It has been shown that internal coordinate systems, when transformed to the modal frame, satisfy the equipartition theorem \cite{Jain2012}.
With, again, the eigenvectors, $\mathbf{\hat{q}}_{i\beta}$ of the generalised inertia $\boldsymbol{\mathcal{I}}_i$ as columns of $\mathbf{Q}_i$, the generalised velocity in the modal frame is given by $\mathbf{Q}_i^T\dot{\boldsymbol{\theta}}_i$, and the modal inertia is $\mathbf{Q}_i^T\mathbf{J}_i^T\boldsymbol{\mathcal{M}_i}\mathbf{J}_i\mathbf{Q}_i$, where $\boldsymbol{\mathcal{M}}_i$ is the mass matrix in Eqn. \ref{eqn:semi-rigid-ke}.
As in previously considered examples, self-consistency dictates that the DoF associated with the motion of atom $j$ in a particular mode is given by its contribution to the modal inertia, which may be calculated for mode $\beta$ by
\begin{equation}
d_{ij}^{\beta} = \frac{\mathbf{\hat{q}}_{i\beta}^T\mathbf{J}_i^T\boldsymbol{\mathcal{M}}_{ij}\mathbf{J}_i\mathbf{\hat{q}}_{i\beta}}{\mathbf{\hat{q}}_{i\beta}^T\mathbf{J}_i^T\boldsymbol{\mathcal{M}}_{i}\mathbf{J}_i\mathbf{\hat{q}}_{i\beta}} = \frac{\mathbf{\hat{q}}_{i\beta}^T\mathbf{J}_i^T\boldsymbol{\mathcal{M}}_{ij}\mathbf{J}_i\mathbf{\hat{q}}_{i\beta}}{\lambda_{i\beta}}, 
    \label{eqn:semi-rigid-df}
\end{equation}
where $\boldsymbol{\mathcal{M}}_i = \sum_{j\in\mathcal{F}_i} \boldsymbol{\mathcal{M}}_{ij}$.
The per-atom mass matrix, $\boldsymbol{\mathcal{M}}_{ij}$, is equivalent to applying a switch matrix, $\mathbf{S}_{ij}$ to $\boldsymbol{\mathcal{M}}_i$, where $\mathbf{S}_{ij}$ contains all zeroes except for the diagonal elements corresponding to atom $j$.
Similar switch matrices can be constructed to obtain directional DoF, or the DoF of a group of atoms within the fragment.
Note, for these semi-rigid fragments, the matrix $\mathbf{J}_i$ depends on the particular instantaneous configuration, and hence the DoF of a given atom may change with time during a simulation, as was noted by Matsubara \textit{et al.} \cite{Matsubara2023}.

To illustrate the process for constructing the Jacobian, consider first the simple example of a water molecule with rigid bond lengths, but an H-O-H angle which is free to change.
Arbitrarily choosing one hydrogen atom (H1) as the reference atom, the Jacobian describing the motion of this semi-rigid fragment is
\begin{equation}
    \bm{J}_{\text{H}_2\text{O}} = \left[\begin{array}{ccc}
        \bm{1} & \bm{0} & \bm{0} \\
        \bm{1} & -\left[\bm{r}_{\text{O}} - \bm{r}_{\text{H1}}\right]_\times & \bm{0} \\
        \bm{1} & -\left[\bm{r}_{\text{O}} - \bm{r}_{\text{H1}}\right]_\times & -\left[\bm{r}_{\text{H2}}-\bm{r}_{\text{O}}\right]_\times
    \end{array}\right],
    \label{eqn:constructing-jacobian}
\end{equation}
where the first block row describes the motion of H1, the second the motion of O, and the third the motion of the second hydrogen, H2.
The first block column accounts for translational motion, while the other two account for rotation of O around H1, and rotation of the O-H2 bond relative to the rotation of O.
Note that other choices are also possible, but any choice which enforces the constraints will produce equivalent results since the inertia matrix is diagonalized.
Also note that omission of the third block column would result in a Jacobian suitable for treating a rigid water molecule, and yields results identical to those in Section \ref{sec:theory:rigid}.

More generally, consider an acyclic (open chain) molecular fragment with atoms connected by a network of rigid bonds.
In this case there is a unique path between any two atoms (tree topology) and it is simple to see that the Jacobian can be constructed in a similar manner, regardless of the number of atoms.
For example, a methyl group with rigid C-H bonds, choosing C as the reference atom, would have the Jacobian
\begin{equation}
    \bm{J}_{\text{CH}_3} = \left[\begin{array}{cccc}
        \bm{1} & \bm{0} & \bm{0} & \bm{0} \\
        \bm{1} & -\left[\bm{r}_{\text{H1}} - \bm{r}_{\text{C}}\right]_\times & \bm{0} & \bm{0} \\
        \bm{1} & \bm{0} & -\left[\bm{r}_{\text{H2}} - \bm{r}_{\text{C}}\right]_\times & \bm{0} \\
        \bm{1} & \bm{0} & \bm{0} & -\left[\bm{r}_{\text{H3}} - \bm{r}_{\text{C}}\right]_\times \\
    \end{array}\right],
\end{equation}
where the block rows correspond to C, H1, H2, and H3, respectively.
This is applicable to the CH$_3$ groups in an ethane molecule where only the C-H bond lengths are held fixed, and we test this system in Section \ref{sec:results:semirigid}.

Constraints other than rigid bond lengths and on fragments with kinematic loops (e.g. in cyclic molecules) require a more complicated process to construct the Jacobian, but can ultimately still be treated under the same framework.
We leave further discussion and application of this as future work.
\vspace{1cm}

It is important to highlight that although partial and local temperature measurements are in principle suitable for use with a thermostat, doing so requires that (1) the thermostat should only be applied to the motion contributing to the temperature measurement, and (2) for constrained molecules undergoing flow, all streaming modal motion must be subtracted before measuring the kinetic temperature.
Point (2) is particularly important; even if kinetic temperature is measured only in directions perpendicular to the flow, rotational motion (e.g. under shear flow) can contribute to the velocities being measured, and hence the streaming angular momentum must still be accounted for.
Similarly, if some mode in a semi-rigid fragment has a non-zero average modal velocity, that too should be treated as streaming motion.

\section{Simulations}

Various molecular dynamics simulations used to verify the above theory are detailed below.
Simulations were performed using LAMMPS \cite{LAMMPS} (version June 2023), with initial configurations generated using Moltemplate \cite{Jewett2021}, Visual Molecular Dynamics (VMD) \cite{VMD} and ``make graphitics'' \cite{make-graphitics_zenodo} as appropriate.
Integration time steps were chosen to ensure that the drift in energy was insignificant using the symplectic velocity Verlet integrator, in which energy corresponding to a shadow Hamiltonian is conserved for step sizes within the stable regime  \cite{Leimkuhler1994,Bond2007,Skeel1999}.
Visualisations were generated using VMD with the Tachyon renderer \cite{TACHYON}.

\subsection{Temperature profile across an interface\label{sec:results:rigid-ethane}}

\begin{figure}[ht]
    \centering
    \includegraphics{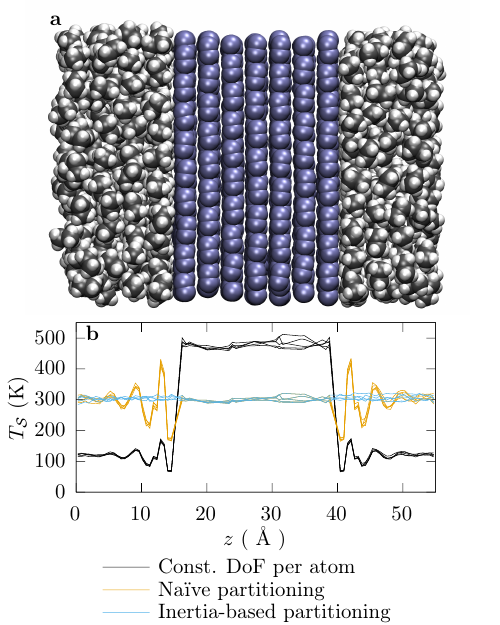}
    \caption{
        Test of local DoF calculation using rigid ethane molecules (white and gray) in contact with a flexible graphene membrane (purple) at equilibrium.
        A snapshot of the system is depicted in (a), aligned with the $x$ axis of the plot in (b) which shows the local temperature of 100 equally sized slab subvolumes centered around the corresponding $z$ values.
        The ``const. DoF per atom'' method calculates local temperature based on the assumption that each atom (whether it is part of a rigid molecule or not) has the same number of DoF associated with it.
        The ``na\"ive'' method assigns the same DoF value to each atom in a rigid body, giving atoms in graphene 3 DoF each, while those in rigid ethane are assigned 6/8 DoF.
        The inertia-based partitioning method uses each atom's contribution to the inertia of each mode, as per Eqn. \ref{eqn:dof-per-atom-total}, resulting in 3 DoF per graphene carbon atom, 1.754 per carbon in rigid ethane, and 0.415 per hydrogen.
        Each line in (b) represents the average temperature profile over a 20 ps time window.
        Note, some bins between graphene sheets had poor statistics due to low occupancy, and were therefore omitted.
    }
    \label{fig:ethane-graphene}
\end{figure}

As a simple prototypical inhomogeneous test system, we model 500 rigid ethane molecules in contact with a flexible 7-layer graphene membrane inside a $3.92\times3.82\times5.5$ nm$^3$ periodic unit cell (see Fig. \ref{fig:ethane-graphene}a).
This gives a system with two regions having approximately equal number density of atoms, but with all rigid constraints concentrated in the ethane phase.
The rigid body equations of motion were integrated using LAMMPS' \verb|rigid| integrator.
The ethane geometry, interaction parameters, and partial charges were from the Automated Topology Builder \cite{Malde2011,Stroet2018} based on the GROMOS 54A7 forcefield \cite{Schmid2011}, and geometric mixing rules were applied for Lennard-Jones parameters between ethane and graphene.
To equilibrate the system at 300 K, we initially applied a Nos\'e-Hoover thermostat \cite{Hoover1985} to the atoms in the graphene membrane and a separate Nos\'e-Hoover thermostat to the rigid ethane molecules, both with a coupling time of 100 fs.
After 50 ps with a 0.5 fs integration time step, the thermostats were removed and the system was further equilibrated for an additional 10 ps under Newton's equations of motion before measuring the kinetic energy in 100 slabs equally spaced in the $z$ direction (perpendicular to the membrane) over a period of 100 ps.
To keep temperature profile data comparable over time, the graphene membrane was restrained in the $z$ direction by coupling its center of mass to a harmonic oscillator with spring constant 5000 kcal mol$^{-1}$\textup{~\AA}$^{-2}$ at the mid-point of the periodic unit cell.
This was done consistently throughout the equilibration and production stages of the simulation.

The total number of DoF throughout the system is three for each graphene atom (total 4032 atoms) plus six per rigid ethane molecule (total 4000 atoms in 500 molecules), or 15096 DoF in total.
Figure \ref{fig:ethane-graphene}b compares the kinetic temperature profiles using three DoF partitioning methods, namely 

\begin{enumerate}
    \item Homogeneous partitioning: dividing 15096 DoF by 8032 atoms and assigning the resultant 1.88 DoF to each atom.
    \item Na\"ive partitioning: assigning 3 DoF to each atom in graphene, and dividing the 6 rigid body DoF evenly between each atom in an ethane molecule (thus 0.75 DoF per ethane C or H).
    \item Inertia-based partitioning: using equation \ref{eqn:dof-per-atom-total} and the rigid ethane geometry gives 1.754 DoF per ethane C and 0.415 DoF per ethane H.
\end{enumerate}
Note that while homogeneous partitioning may seem overly simplistic, it is the default in some molecular dynamics packages, and a reasonable approach for homogeneous systems when local temperatures are measured within reasonably sized subvolumes.

The inertia-based DoF partitioning scheme yields a uniform kinetic temperature profile of 300 K as expected, despite the system's inhomogeneity, while the na\"ive method results in artificial ringing at the interface due to the packing of ethane molecules against the flat surface.
As the na\"ive method underestimates the DoF of C and overestimates that of H, this corresponds to relatively low apparent temperature in regions with high H concentration, and relatively high apparent temperature in regions with high concentration of ethane C atoms.
The homogeneous method, predictably, yields significant error throughout the system due to underestimating the DoF of atoms in graphene, and overestimating that of atoms in rigid ethane.

\subsection{Directional temperature\label{sec:results:water}}

\begin{figure*}[ht]
    \centering
    \includegraphics{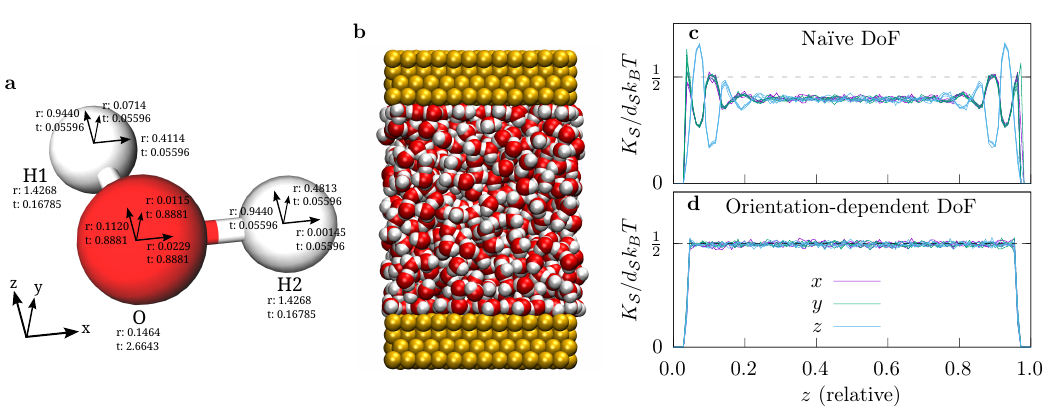}
    \caption{%
    Directional DoF calculation and local temperature measurement of rigid SPC/E \cite{Berendsen1987} water molecules.
    (a) Example results for the rotational, r, and translational, t, DoF in each Cartesian direction (on axes) and in total (below atom labels) for a water molecule lying in the $xy$-plane with the $x$ axis aligned with the O--H2 bond.
    (b) Illustrates the test system, (c) shows the directional kinetic energy of hydrogen atoms in each subvolume, $K_\mathcal{S}$, normalised by a na\"ive approximation of the DoF in each subvolume calculated by dividing the 6 rigid body DoF evenly between each direction of each atom (i.e. 2/3 each in each direction), and (d) shows the same normalised instead by the orientation-dependent directional DoF, as calculated by Eqn. \ref{eqn:dof-directional}.
    Dashed lines in (c) and (d) indicate the expected value of $0.5k_BT$ per DoF, and four sequential 100 ps averages have been superimposed for each observable.
    }
    \label{fig:confined-water}
\end{figure*}

 Results for an example calculation of the directional DoF of atoms in a rigid water molecule are shown in Fig. \ref{fig:confined-water}a.  
The water molecule is oriented so that it lies in the $xy$-plane, with one O--H bond along the $x$ axis of the Cartesian frame.
Note, although the two hydrogen atoms are symmetrically equivalent and their total rotational and translational DoF are the same, the directional rotational DoF components are different.
This is because the reference frame of the calculation does not correspond to the modal frame (i.e. the reference frame's Cartesian axes do not align with the principal axes of the rigid body).

We test this directional DoF assignment by simulating 1000 rigid water molecules confined between two walls, as illustrated in Fig. \ref{fig:confined-water}b, and analysing the local, directional kinetic temperature of the hydrogen atoms. 
The 1152 wall particles were modelled as copper atoms (mass 63.546 amu) on a FCC lattice with (100) surface.
Lennard-Jones interaction parameters for Cu were as for Ref. \cite{Heinz2008}, with arithmetic mixing rules for Cu-O interactions.
Rigid water molecules were of the simple point charge extended (SPC/E) type \cite{Berendsen1987}, and were integrated using Newton's equations of motion.
For initialization and initial equilibration, using an integration time step of 1 fs, the initial configuration was allowed to relax for 2 ns with the inter-wall spacing slowly adjusted from 7.25 nm until the density in the central 20\% of the channel matched that of a constant pressure simulation of SPC/E water at the same temperature and 1 atm.
During this process, the wall atoms were fixed at their lattice points, while a Nos\'e-Hoover thermostat at 300 K with $\tau = 100$ fs was applied to the water molecules.
The simulation box size was fixed throughout at $3.063 \times 2.6897 \times 6.8$ nm$^3$, and the final channel width between surface wall particle centers was approximately 3.79 nm.
The system was then equilibrated with fixed inter-wall distance for a further 4 ns, before reducing the integration time step to 0.5 fs and equilibrating for another 100 ps.
Each wall particle was restrained about its initial lattice position with a three-dimensional harmonic potential of spring constant $k = 65$ kcal mol$^{-1}$\textup{~\AA}$^{-2}$), or a fundamental oscillatory period of $\sim48$ fs.
A three-particle chain Nos\'e-Hoover thermostat was applied to the combined group of upper and lower wall particles with temperature $T=300$ K and coupling time $\tau = 100$ fs, using the LAMMPS \verb|tloop 5| keyword to integrate the thermostat variables over five substeps per main integration step for added accuracy.
SHAKE \cite{Ryckaert1977}, RATTLE \cite{Andersen1983} (both with $10^{-4}$ relative accuracy), or LAMMPS' \verb|rigid| integrator were used to impose constraints, with differences between constraint algorithms all within statistical uncertainty.
The data presented in Fig. \ref{fig:confined-water} used the SHAKE algorithm.

Figure \ref{fig:confined-water}c shows that due to ordering of the rigid water molecules near the walls, kinetic energy is not equally partitioned between Cartesian directions.
Considering that the system is at equilibrium, equipartition of kinetic energy between the degrees of freedom is expected.
Where the the water molecules are randomly oriented, like in the bulk, the average orientation-dependent directional components of the rotational degrees of freedom would average out.
However, this is not the case near the interface, and it is clear that orientational ordering of the molecules at the interface causes some directions to have different DoF to others.
In Fig. \ref{fig:confined-water}d, we introduce a scaling factor equal to the total DoF in the given local volume for motion in the given direction, which is dynamically calculated based on the orientation of each water molecule.
This clearly shows that $\left<T_{\mathcal{S},\gamma}\right> = T$, $\forall$ $\gamma\in\mathcal{D}$, $\mathcal{S}\subseteq\mathcal{P}$; that is, the ensemble average of the local, directional temperature agrees with the set-point temperature for all local subsets and all Cartesian directions, as expected at equilibrium.

\subsection{Temperature gradient across rigid bodies\label{sec:results:dumbbells}}

\begin{figure}[ht]
    \centering
    \includegraphics{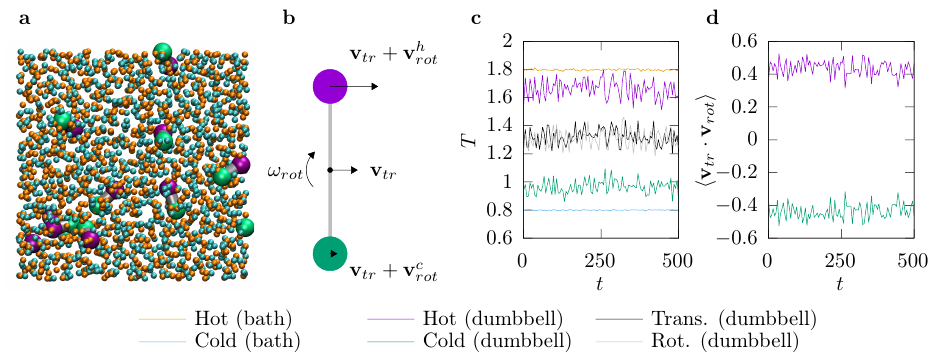}
    \caption{%
        Demonstration of temperature gradient across rigid dumbbells.
        Panel (a) shows a snapshot of the molecular dynamics simulation.
        The purple (green) particle of each dumbbell interacts only with particles of the same color and with the orange (cyan) bath particles.
        Orange bath particles are thermostatted to a normalised temperature of 1.8, and cyan particles a temperature of 0.8.
        All particles are the same size and mass, although dumbbells are drawn larger for clarity.
        (b) Illustrates a situation in which one end of a dumbbell carries more kinetic energy than the other, despite excitations of the rotational and translational modes being of similar magnitude.
        (c) Shows the kinetic temperature of bath and dumbbell particles, and of the translational and rotational dumbbell modes, plotted with time at the steady state (averaged over bins of 5 LJ time units for clarity).
        Similarly, (d) shows the correlation between the translational and rotational velocity components of the hot and cold dumbbell particles, averaged over the 10 dumbbells in the system.
    }
    \label{fig:dumbbell-bath}
\end{figure}

In simulations of long molecules where there is a temperature gradient present, it is likely that different ends of the molecule will be in different thermal environments, and the approach developed here should enable examination of this for molecules with constraints.
A simple test for a temperature gradient across a rigid body can be constructed using rigid dumbbells consisting of two point particles; one interacting with a bath of hot particles, and one with a bath of cold particles, as illustrated in Fig. \ref{fig:dumbbell-bath}a.
The hot and cold bath particles occupy the same physical space but do not interact directly, hence heat transfer must occur through the rigid dumbbells.
All particles, including those that formed the rigid dumbbells, were simple Lennard-Jones (LJ) particles of unit size and mass with an interaction cut-off of 2.6 LJ units.
The hot and cold particles were coupled to separate Nos\'e-Hoover thermostats at $T_H=1.8$ and $T_C=0.8$, respectively, with a coupling time of $0.1$, all in normalised LJ units.
The LAMMPS \verb|rigid| integrator was used to maintain the rigid constraints, and equations of motion were integrated with a dimensionless time step of 0.001.
The cubic periodic unit cell of side length 10.77 LJ units contained 990 hot particles, 990 cold particles, and 10 dumbbells with bond length 1.077, for a total effective density of $0.8$ for each set of interacting particles.
This ensured that both baths were in the liquid phase.
Free particles simply have 3 DoF each, while each particle within a dumbbell has 2.5 DoF (1.5 translational and 1 rotational).
Particles were initially generated on a cubic lattice, and then simulated for 250 LJ time units to reach the steady state before gathering data.

Figure \ref{fig:dumbbell-bath}c shows the kinetic temperatures of the baths, the hot and cold ends of the dumbbells, and the translational and rotational modes of the dumbbells in the steady state.
It can be seen that while the overall translational and rotational temperatures of the dumbbells are approximately equal, the dumbbell particles in contact with the hot bath have significantly more kinetic energy on average than the ones in contact with the cold bath.
This implies some correlation between the translational and rotational velocity components, resulting in a situation which is on average similar to that illustrated in Fig. \ref{fig:dumbbell-bath}b.
Figure \ref{fig:dumbbell-bath}d plots the correlation function between the velocity component of a particle in a rigid dumbbell due to translation and that due to rotation, clearly showing a positive correlation for the hot particle, and a negative one for the cold particle.
Thus, the rigid body motion is such that one end of the dumbbell is consistently and significantly hotter than the other.

\subsection{Semi-rigid Fragments\label{sec:results:semirigid}}

We test the calculation for DoF of semi-rigid fragments again on ethane, this time in the bulk phase, with three directional temperatures calculated from the set of all C atoms, and another three from the set of all H atoms.
In this case, the C--H bond lengths are fixed, but bond angles, dihedral angles, and C--C bond lengths are free to change, as is common in molecular dynamics force fields for organic molecules.
1000 ethane molecules, initially placed on a $10\times10\times10$ grid, were simulated in a $7\times7\times7$ nm$^3$ periodic volume, with interaction parameters as for Section \ref{sec:results:rigid-ethane} and an integration time step of 0.5 fs.
Atoms were given initial velocities from a Boltzmann distribution at 300 K, and allowed to relax under a Nos\'e-Hoover chain thermostat at $T=300$ K with a 100 fs coupling time and a chain length of 3.
C--H bond lengths were fixed using the RATTLE \cite{Andersen1983} algorithm with a $10^{-4}$ relative accuracy.
A $10^{-5}$ relative accuracy was also trialed, and found to have negligible effect on results.

Table \ref{tab:ethane-dof-dist} shows that during the simulation, C and H atoms have geometry-dependent DoF that are almost always similar to values from an energy-minimized geometry.
The resulting DoF (about 2.77 per C atom and 2.08 per H atom) differ significantly from a uniform distribution of 2.25 DoF per atom (24 unconstrained DoF, minus 6 C-H bond length constraints, divided by 8 atoms).
Based on uniformly-distributed DoF, the trajectory-averaged local kinetic temperatures of C and H atoms are 374 K and 276 K respectively and give no meaningful information about thermodynamic observables.

\begin{figure}[ht]
    \centering
    \includegraphics{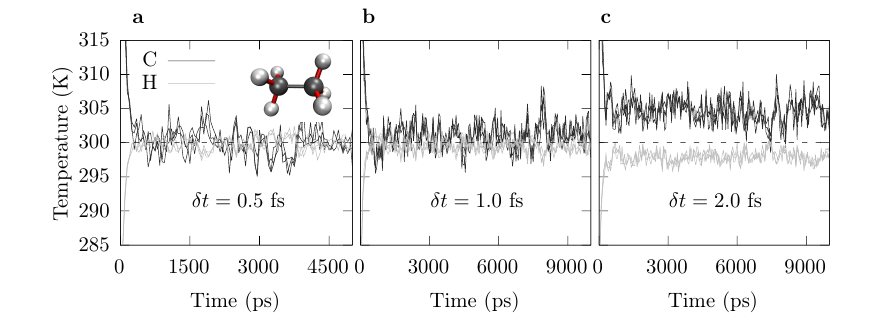}
    \caption{%
    Directional kinetic temperature of hydrogen and carbon atoms in ethane with rigid C--H bond lengths (highlighted red in the inset), using an integration time step of (a) 0.5 fs, (b) 1.0 fs, and (c) 2.0 fs.
    The $x$, $y$, and $z$ components of the temperature are plotted as separate lines,  showing the average value over all time steps within a 50 ps window for clarity.
    The system was thermostatted to 300 K, with randomly sampled initial velocities and molecules initially positioned on a uniform lattice.
    }
    \label{fig:ethane-vs-timestep}
\end{figure}

Figure \ref{fig:ethane-vs-timestep} shows by contrast that inertia-based DoF partitioning gives meaningful local kinetic temperatures.
At the smallest integration time step shown, 0.5 fs, the local kinetic temperatures of both C and H atoms quickly converge to 300 K as the long-time expectation value underlying instantaneous fluctuations, in agreement with the set-point of the thermostat and the temperature of the overall system.
Thus, it is clear that the inertia-based DoF partitioning yields a correct local temperature measurement.
However, Figure \ref{fig:ethane-vs-timestep} also shows that at larger time steps the C and H atoms converge to different temperatures whose difference from the expected temperature scales quadratically with the time step (see Fig. S1 in the Supporting Information), reaching 4.5 K at 2.0 fs (commonly used in biomolecular MD simulations).
Notably, this is despite the system-wide global kinetic temperature converging to 300 K within 50 ps in all cases.
It has previously been shown that, for the integration scheme used in this work, rigid constraints introduce a discretization error in the configurational distribution which scales with the square of the time step \cite{Zhang2019}, as is observed in Fig. S2 of the Supporting Information.
The probability distributions of structural features (bond lengths, angles and dihedrals) are also affected by the discretization error, with a slight broadening at larger time steps indicative of the configurational DoF being too hot, as shown in Fig. S3-S5 of the Supporting Information, .
Hence, we expect these temperature gaps are indicative of similar discretization errors in the local kinetic energy distributions, which cancel out when only the total kinetic energy is measured.
Similar issues have previously been highlighted for rigid water molecules \cite{Asthagiri2023}.

Thus, beyond their fundamental importance to nonequilibrium properties (such as temperature gradients) and dynamic properties (such as diffusivities and other transport coefficients), local temperatures (e.g. grouped by symmetrically equivalent atoms) may additionally serve as a useful indicator for the onset of discretization error in integration schemes which otherwise preserve the distribution of the total kinetic energy.
Importantly, checking for discretization error using local atomic temperatures requires only structure and momentum data, and does not require additional simulations with smaller time steps as a comparison point, thereby presenting a lower computational cost than analysis of potential energy.
We note, however, that because force fields are parameterized using a particular time step, any time step-dependent discretization error may be implicitly accounted for in parameterization, and hence reducing discretization error may not correspond with improved agreement between simulated and experimental results.

\begin{table}[h]
\caption{%
        Distribution of the DoF for the H and C atoms in ethane with rigid C--H bond lengths.$^a$
        }
    \vspace*{0.2cm}
    \centering
    \begin{tabular}{c|c|c|c|c}
        Atom & Min. DoF & Max. DoF & Mean DoF$\hspace{1.5pt}^b$ & Std. Dev. \\\hline
        C & 2.7685 & 2.7742 & 2.7709 & $5.8\times10^{-4}$ \\
        H & 2.0746 & 2.0773 & 2.0764 & $2.8\times10^{-4}$
    \end{tabular} 
    \parbox{11.5cm} {\footnotesize \begin{flushleft}
    $^a$ From simulations at equilibrium (300 K, 0.5 fs time step).
    
    $^b$ The mean values are in good agreement with those of the initial geometry taken from the ground state of a single molecule energy minimisation calculation.

\end{flushleft}}
    \label{tab:ethane-dof-dist}
\end{table}

\section{Conclusion}
In summary, we have derived a general and self-consistent framework for partitioning the degree of freedom associated with a given mode between participating masses by requiring only that any subset of masses has the same kinetic temperature within the mode.
This leads to the simple result that the fraction of a mode's DoF associated with a particular mass is equal to the inertia of that mass in the modal motion divided by the total inertia of the mode.
This enables both local and directional kinetic temperature measurement, and is applicable both to rigid and semi-rigid fragments consisting of point masses and/or volumetric rigid bodies.
Results were validated on molecular dynamics simulations of inhomogeneous systems, where the utility of the method is evident.
Importantly, this method enables local kinetic temperature measurements which encompass all relevant DoF, and, unlike other methods recently presented in literature, it is general over arbitrarily constrained systems and can provide both modal and directional information.
Finally, by measuring separately the temperatures of carbon and hydrogen atoms with commonly-used rigid C-H bonds, we have shown that for integration time steps above 0.5 fs, a steady state is reached in which the two atom types have divergent temperatures, which may have implications on the development and transferability of molecular dynamics force fields.\\

{\bf SUPPORTING INFORMATION}: Analysis of the potential energy distribution (PDF)

\begin{acknowledgments}
The authors thank the Australian Research Council for its support for this project through the Discovery program (FL190100080).
We acknowledge access to computational resources provided by the Pawsey Supercomputing Centre with funding from the Australian Government and the government of Western Australia, and the National Computational Infrastructure (NCI Australia), an NCRIS enabled capability supported by the Australian Government.
We also thank A/Prof. Taras Plakhotnik for posing an interesting problem which led to the inspiration for this work.
\end{acknowledgments}

\appendix
\bibliographystyle{achemso}
\bibliography{references_abbrev}

\end{document}



\title{Supporting Information: \\ Local temperature measurement in molecular dynamics simulations with rigid constraints}

\author{Stephen Sanderson}
 \email{stephen.sanderson@uq.edu.au}
 \affiliation{Australian Institute for Bioengineering and Nanotechnology, The University of Queensland, St. Lucia, QLD, 4072, Australia}
\author{Shern R. Tee}
 \affiliation{Australian Institute for Bioengineering and Nanotechnology, The University of Queensland, St. Lucia, QLD, 4072, Australia}
 \affiliation{School of Environment and Science - Chemistry and Forensic Science, Griffith University, Nathan, QLD, 4111, Australia}
\author{Debra J. Searles}
 \email{d.bernhardt@uq.edu.au}
 \affiliation{Australian Institute for Bioengineering and Nanotechnology, The University of Queensland, St. Lucia, QLD, 4072, Australia}
 \affiliation{School of Chemistry and Molecular Biosciences, The University of Queensland, St. Lucia, QLD, 4072, Australia }
  \affiliation{ARC Centre of Excellence for Green Electrochemical Transformation of Carbon Dioxide, The University of Queensland, St. Lucia, QLD, 4072, Australia }

\date{\today}

\maketitle
\renewcommand{\thesection}{S\arabic{section}}
\renewcommand{\thefigure}{S\arabic{figure}}
\renewcommand{\theequation}{S\arabic{equation}}

\section{Analysis of the potential energy distribution}

\begin{figure}
    \centering
    \includegraphics[]{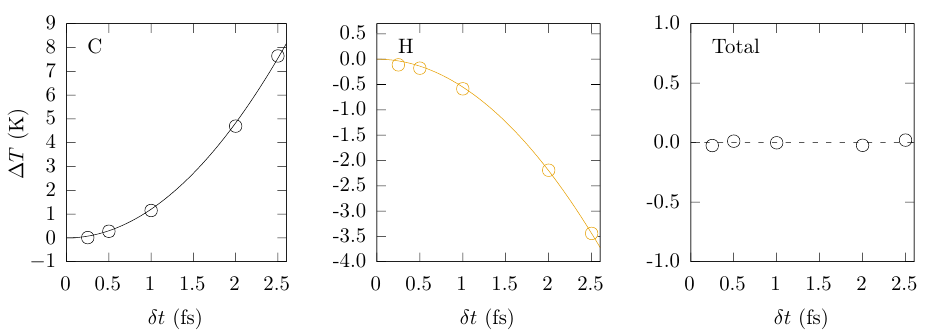}
    \caption{Average difference between temperature of C atoms (left), H atoms (middle), and all atoms (right), and the temperature set by the thermostat  (300 K) for ethane with rigid C--H bonds.
    The solid line shows a fit of the form $a\times\delta t^2$.
    Averages were taken over a 3900 ps period after 400 ps of equilibration time.
    }
    \label{fig:ethane-T-vs-dt}
\end{figure}

As shown in Fig. \ref{fig:ethane-T-vs-dt}, the difference in temperature between C and H atoms and the thermostat set-point appears to increase quadratically with the size of the integration time step.
This observation is despite the fact that the total kinetic energy of the system agrees with the set-point of 300 K.
Hence, it is apparent that, at least for the simulation conditions and integrator used, longer time steps result in a steady state in which C atoms are hotter than H atoms.

\begin{figure}
    \centering
    \includegraphics[]{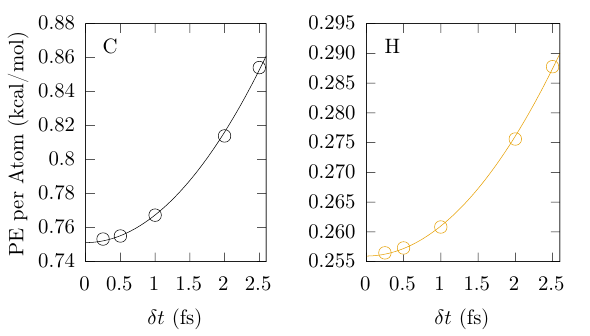}
    \caption{Average potential energy per atom of C (left) and H (right) atoms in ethane with rigid C--H bonds for varying integration time steps.
    The solid line shows a fit of the form $a\times\delta t^2+b$.
    All data averaged over 3900 ps, after a 400 ps equilibration time.
    }
    \label{fig:ethane-pe-vs-dt}
\end{figure}

\begin{figure}
    \centering
    \includegraphics[]{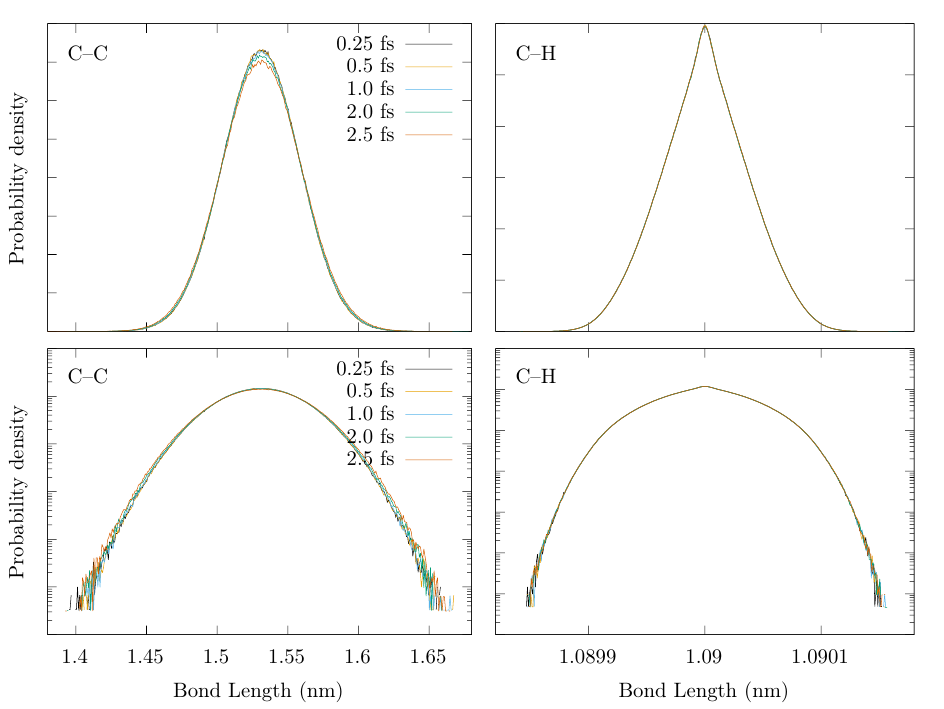}
    \caption{Bond length distributions of flexible C--C (left) and rigid C--H (right) bonds in ethane for various integration time steps, shown on a linear (top) and logarithmic (bottom) scale.
    Distributions measured over 3900 ps using snapshots 1 ps apart, beginning after 400 ps of eqiuilibration time.
    }
    \label{fig:ethane-bond-dist}
\end{figure}

\begin{figure}
    \centering
    \includegraphics[]{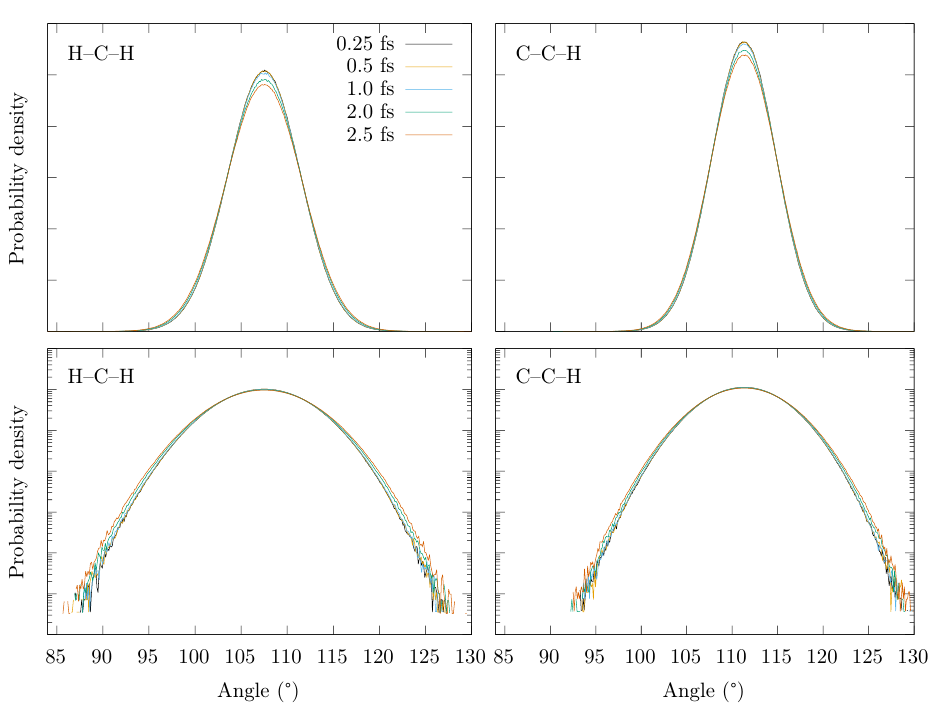}
    \caption{Bond angle distributions of H--C--H (left) and C--C--H (right) angles in ethane with rigid C--H bonds for various integration time steps, shown on a linear (top) and logarithmic (bottom) scale.
    Distributions measured over 3900 ps using snapshots 1 ps apart, beginning after 400 ps of eqiuilibration time.
    }
    \label{fig:ethane-angle-dist}
\end{figure}

\begin{figure}
    \centering
    \includegraphics[]{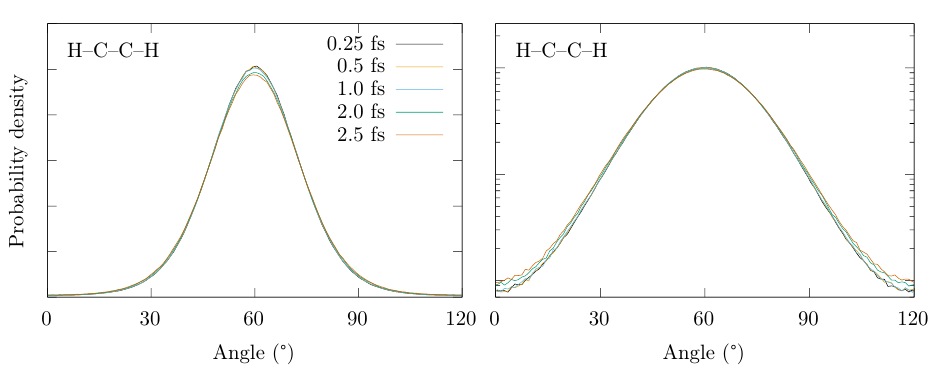}
    \caption{Distributions of the H--C--C--H dihedral angle in ethane with rigid C--H bonds for various integration time steps, shown on a linear (left) and logarithmic (right) scale.
    Distributions measured over 3900 ps using snapshots 1 ps apart, beginning after 400 ps of eqiuilibration time.
    Measurements were taken only from the single dihedral angle of each molecule which had a non-zero force constant in the force field, and averaged over the 0--120\textdegree, 120--240\textdegree, and 240--360\textdegree ranges to reduce noise.
    Hence, only the 0--120\textdegree range is shown for clarity.
    }
    \label{fig:ethane-dihedral-dist}
\end{figure}

The velocity Verlet integrator, as implemented in LAMMPS, is expected to preserve the kinetic energy distribution better than the potential energy distribution \cite{Zhang2019}, and Fig. \ref{fig:ethane-pe-vs-dt} indeed shows that the potential energy also scales approximately quadratically with the size of the time step.
Furthermore, Figures \ref{fig:ethane-bond-dist}--\ref{fig:ethane-dihedral-dist} demonstrate that while the error in the rigid bond lengths remains the same at larger time steps, the bond length, bond angle, and dihedral angle distributions broaden slightly, indicating overheating of the configurational degrees of freedom.
These results indicate that while the integrator may preserve the distribution of the total system kinetic energy out to relatively long time steps, it may not preserve the local details of that distribution.
It is therefore suggested that, at least for integrators which accurately preserve the momentum distribution, breaking of equipartition between local kinetic temperatures in an equilibrium simulation may be a good indicator of reduced accuracy in the configurational distribution.
Importantly, this metric does not require computationally expensive comparisons against other integration time steps.

\bibliographystyle{achemso}
\bibliography{references_abbrev}